\documentclass[preprint,twocolumn]{aastex63}
\usepackage{amsmath}
\shorttitle{Primordial graviton background and inflation}
\shortauthors{Vagnozzi \& Loeb}

\begin{document}

\title{The challenge of ruling out inflation via the primordial graviton background}

\author[0000-0002-7614-6677]{Sunny Vagnozzi}
\affiliation{Department of Physics, University of Trento, Via Sommarive 14, 38123 Povo (TN), Italy}
\affiliation{Kavli Institute for Cosmology, University of Cambridge, Madingley Road, Cambridge CB3 0HA, United Kingdom}
\correspondingauthor{Sunny Vagnozzi}
\email{sunny.vagnozzi@unitn.it}

\author[0000-0003-4330-287X]{Abraham Loeb}
\affiliation{Department of Astronomy, Harvard University, 60 Garden Street, Cambridge, MA 02138, USA}

\begin{abstract}

\noindent Recent debates around the testability of the inflationary paradigm raise the question of how to model-independently discriminate it from competing scenarios. We argue that a detection of the Cosmic Graviton Background (CGB), the relic radiation from gravitons decoupling around the Planck time, would rule out the inflationary paradigm, as realistic inflationary models would dilute the CGB to an unobservable level. The CGB contribution to the effective number of relativistic species, $\Delta N_{{\rm eff},g} \approx 0.054$, is well within the reach of next-generation cosmological probes. We argue that detecting the high-frequency stochastic gravitational wave background associated to the CGB will be challenging but potentially feasible. We briefly discuss expectations within alternatives to inflation, focusing on bouncing cosmologies and emergent scenarios.

\end{abstract}


\section{Introduction}
\label{sec:intro}

Inflation, a postulated stage of quasi-de Sitter expansion in the primordial Universe, is widely regarded as the leading paradigm for the very early Universe. Originally introduced to address various fine-tuning problems of the hot Big Bang (hBB) model, inflation provides a compelling mechanism for generating the density perturbations from which structure eventually originated~\citep{Starobinsky:1980te,Guth:1980zm,Mukhanov:1981xt,Linde:1981mu,Albrecht:1982wi}. The predictions of some of the simplest inflationary models are in remarkable agreement with observations of the Cosmic Microwave Background (CMB) and the Large-Scale Structure (LSS), which in turn is commonly viewed as a sign of the inflationary paradigm's success.

Despite these successes, inflation is not free of open issues, and over the years criticisms have been raised about its status~\citep[see e.g.][]{Ijjas:2014nta,Martin:2019zia}. One of the major bones of contention is driven by the large flexibility with regards to the predictions of individual inflationary models, and concerns whether or not the inflationary paradigm is falsifiable. We use the term ``paradigm'' and not ``model'' since any given inflationary model is clearly falsifiable, whereas these doubts concern the inflationary scenario as a whole.  Here we do not seek to take sides in the debate, but simply note that these issues strongly motivate the question of how to \textit{model-independently} discriminate the inflationary paradigm from alternative scenarios for the production of density perturbations.

We address the above question by identifying a signature \textit{de facto} precluded to any realistic inflationary model, and whose observation would thus rule out the inflationary paradigm. The decoupling of primordial gravitons around the Planck time should leave behind a thermal background of relic gravitons: the Cosmic Graviton Background (CGB). An inflationary phase taking place between the Planck era and today would wash out the CGB, rendering it unobservable: an unambiguous CGB detection would therefore pose a major threat to the inflationary paradigm. In this \textit{Letter}, we formalize these arguments and discuss prospects for detecting the CGB.

\section{The Cosmic Graviton Background}
\label{sec:cgb}

We now discuss the features of the CGB in the absence of inflation. We adopt the working assumption that above the Planck scale point-like four-particle interactions involving two gravitons, whose rate at temperature $T$ is of order $\Gamma_g \sim T^5/M_{\rm Pl}^4$, kept gravitons in thermal equilibrium in the primordial plasma~\citep[see also][]{Zhao:2009pt,Giovannini:2019oii}. If we assume adiabatic evolution throughout the early stages of the primordial plasma, and therefore that the Universe was radiation dominated up to then, the Hubble rate scales as $H \sim T^2/M_{\rm Pl}$. Comparing the two rates indicates that gravitons decouple at a temperature $T_{g,{\rm dec}} \sim M_{\rm Pl}$ (or equivalently around the Planck time $t_{g,{\rm dec}} \sim t_{\rm Pl}$): besides ruling out inflation, a detection of the CGB would thus provide an experimental testbed for theories attempting to unify quantum mechanics and gravity.

Being massless and thus decoupling while relativistic, primordial gravitons preserve the blackbody form of their spectrum following decoupling, with the effective CGB temperature $T_g$ redshifting with the scale factor $a$ as $T_g \propto 1/a$. Since the entropy density $s=2\pi^2g_{\star}^s(T)T^3/45$ scales as $s \propto a^{-3}$, where $g_{\star}^s(T)$ is the (temperature-dependent) effective number of entropy degrees of freedom (DoF), we can relate the present-day temperatures of the CGB and CMB, $T_{g,0}$ and $T_{\gamma,0}$ respectively, as follows:
\begin{eqnarray}
\frac{T_{g,0}}{T_{\gamma,0}} = \left ( \frac{g_{\star}^s(T_0)}{ \left ( g_{\star}^s(T_{\rm Pl})-2 \right ) } \right )^{1/3}\,,
\label{eq:tg0tgamma0}
\end{eqnarray}
where $g_{\star}^s(T_0) \simeq 3.91$ is the present-day effective number of entropy DoF \textit{excluding} gravitons (accounting for photons and neutrinos), and $g_{\star}^s(T_{\rm Pl})$ is the effective number of entropy DoF prior to graviton decoupling, \textit{including} gravitons. Accounting only for Standard Model (SM) DoF up to the Planck scale, above the electroweak (EW) scale $g_{\star}^s(T_{\rm Pl})-2 \simeq 106.75$. Precise measurements of the CMB frequency spectrum from COBE/FIRAS fix $T_{\gamma,0} \approx 2.7\,{\rm K}$ and therefore under these minimal assumptions the present-day CGB temperature is predicted to be $T_{g,0} \simeq (3.91/106.75)^{1/3}T_{\gamma,0} \approx 0.9\,{\rm K}$, making the CGB about 3 times colder than the CMB.

Lacking a precise knowledge of the type of new physics lying beyond the ${\rm TeV}$ scale, the assumption of only considering SM DoF up to the Planck scale is conservative, but likely somewhat unrealistic, as one might expect several additional DoF to appear in the ``desert'' between the EW and Planck scales. If so, $g_{\star}^s(T_{\rm Pl})$ in the denominator of Eq.~(\ref{eq:tg0tgamma0}) can only increase, decreasing $T_{g,0}$ with respect to the previous estimate $T_{g,0} \approx 0.9\,{\rm K}$, which therefore should be viewed more as a conservative upper bound on $T_{g,0}$. However, the exact numbers are highly model-dependent and depend on the specific new physics model. For instance, $T_{g,0} \approx 0.7\,{\rm K}$ in a supersymmetric-like scenario where $g_{\star}^s(T_{\rm Pl})$ doubles, whereas $T_{g,0} \approx 0.4\,{\rm K}$ in a hypothetical scenario where $g_{\star}^s(T_{\rm Pl})$ increases by an order of magnitude.

\section{Can inflation be ruled out?}
\label{sec:inflation}

Our assumption of adiabatic evolution from $T_{\rm Pl}$ down to present times breaks down whenever comoving entropy is generated, e.g.\ during reheating at the end of inflation. An inflationary phase alters the relation between $T_{g,0}$ and $T_{\gamma,0}$ in Eq.~(\ref{eq:tg0tgamma0}), as the latter would be determined by the dynamics of reheating, which however can at most produce out-of-equilibrium graviton excitations, unless the effective gravitational constant $G_{\rm eff}$ was significantly higher at reheating. Since the scale factor increases exponentially during inflation, the CGB temperature itself is exponentially suppressed by a factor of $e^{-N}$, with $N$ the number of \textit{e}-folds of inflation.

We can obtain an extremely conservative upper limit on $\widetilde{T}_{g,0}$ in the presence of a phase of inflation (the tilde distinguishes the present-day graviton temperatures with and without inflation), using the facts that \textit{a)} solving the horizon and flatness problems requires $N \gtrsim 60$, and \textit{b)} reheating should occur at $T_{\rm rh} \gtrsim 5\,{\rm MeV}$ in order to not spoil Big Bang Nucleosynthesis predictions~\citep{deSalas:2015glj}. From these requirements we find that $\widetilde{T}_{g,0} \lesssim 50\,\mu{\rm K}$, implying that inflation would dilute the CGB to an unobservable level. More generically, we find the following upper limit:
\begin{eqnarray}
\widetilde{T}_{g,0} \ll 0.25 \left ( \frac{T_{\rm rh}}{{\rm GeV}} \right ) ^{-1}e^{60-N}\,\mu{\rm K}\,.
\label{eq:upperlimitinflation}
\end{eqnarray}
However, $\widetilde{T}_{g,0} \lesssim 50\,\mu{\rm K}$ is a very conservative upper limit, for two reasons. Firstly, in most realistic models inflation typically proceeds for more than 60 \textit{e}-folds, leading to further exponential suppression [see Eq.~(\ref{eq:upperlimitinflation})]. Next, although reheating at scales as low as $T_{\rm rh} \simeq {\cal O}({\rm MeV})$ is observationally allowed, models realizing this in practice are very hard to construct~\citep[see e.g.][]{Kawasaki:1999na,Hannestad:2004px,Khoury:2011ii}.~\footnote{Note, however, that a viable interpretation of the signal recently observed by various Pulsar Timing Arrays~\citep[e.g.][]{NANOGrav:2020bcs} is in terms of inflationary GWs given a rather low reheating scale~\citep{Vagnozzi:2020gtf,Kuroyanagi:2020sfw,Oikonomou:2021kql,Odintsov:2021kup,Benetti:2021uea,Oikonomou:2022ijs}.} It if far more likely that, if inflation did occur, reheating took place way above the EW scale, further tightening the upper bound on $T_{g,0}$.

One may try to evade these conclusions invoking models of \textit{incomplete inflation} with a limited number of \textit{e}-folds $46 \lesssim N \lesssim 60$: however, if inflation is indeed the solution to the flatness problem, such models are essentially ruled out by current stringent bounds on spatial curvature~\citep{Vagnozzi:2020dfn}, as argued explicitly in~\cite{Efstathiou:2020wem}. Even if $N<60$, bringing $\widetilde{T}_{g,0}$ to a detectable level still requires an extremely low reheating scale, typically harder to achieve within models of incomplete inflation.

A caveat to our previous results is our assumption of inflation occurring at sub-Planckian scales. Specifically, $T_{\rm rh}>M_{\rm Pl}$ is required for the CGB not to be washed out by inflation. However, on general grounds there are serious concerns about the consistency of trans-Planckian effects both during inflation and at reheating~\citep[e.g.][]{Brandenberger:2012aj,Brandenberger:2022pqo}. A specific concern is given by the trans-Planckian censorship conjecture, which sets tight limits on the maximum inflationary scale $\Lambda_{\rm inf}^{\max}$ and reheating temperature: $\Lambda_{\rm inf}^{\max}\,, T_{\rm rh} \ll M_{\rm Pl}$~\citep{Bedroya:2019snp,Bedroya:2019tba,Mizuno:2019bxy,Kamali:2019gzr}.

More importantly, the lack of detection of inflationary B-modes indicates that $\Lambda_{\rm inf}^{\max}$ is at least four orders of magnitude below the Planck scale. For instantaneous reheating, the reheating temperature is obviously limited to $T_{\rm rh}<\Lambda_{\rm inf}^{\max}$, as reheating to higher temperatures would violate (covariant) stress-energy conservation. For non-instantaneous reheating, $T_{\rm rh}$ is of course even lower~\citep[see also][]{Cook:2015vqa}. Therefore, we deem it very safe to assume that $T_{\rm rh} \ll M_{\rm Pl}$, corroborating all our earlier findings. In summary, within realistic inflationary cosmologies one does not expect to be able to detect the relic thermal graviton background -- conversely, a convincing detection thereof would rule out the inflationary paradigm.

\section{Detectability of the CGB}
\label{sec:detectability}

We now investigate whether detecting the CGB is experimentally feasible, considering our benchmark $T_{g,0} \approx 0.9\,{\rm K}$ case. The contribution of the CGB to the effective number of relativistic species $N_{\rm eff}$ is given by:
\begin{eqnarray}
\Delta N_{{\rm eff},g} \equiv \frac{8}{7} \left ( \frac{11}{4} \right )^{\frac{4}{3}}\frac{\rho_g}{\rho_{\gamma}} = \frac{8}{7} \left ( \frac{11}{4} \right )^{\frac{4}{3}} \left ( \frac{g_{\star}^s(T_0)}{ \left ( g_{\star}^s(T_{\rm Pl})-2 \right ) } \right )^{\frac{4}{3}}\,.
\label{eq:neffg}
\end{eqnarray}
For $g_{\star}^s(T_{\rm Pl})-2=106.75$, we therefore find that $\Delta N_{{\rm eff},g} \approx 0.054$, as expected for a species with 2 spin DoF decoupling before the QCD phase transition.

A contribution to $N_{\rm eff}$ of this size is a factor of $3$ below the sensitivity of current probes. However, this number is well within the reach of a combination of next-generation CMB and LSS surveys. For instance, even after marginalizing over the total neutrino mass, \cite{Brinckmann:2018owf} forecast a sensitivity of $\sigma_{N_{\rm eff}} \simeq 0.021$ combining CMB data from CMB-S4 and LiteBIRD with galaxy clustering and cosmic shear data from Euclid, whereas with a PICO-like experiment in place of CMB-S4+LiteBIRD the sensitivity improves to $\sigma_{N_{\rm eff}} \simeq 0.017$. Therefore, if the benchmark $0.9\,{\rm K}$ CGB were present, CMB-S4+LiteBIRD+Euclid would be able to detect it through its imprint on $N_{\rm eff}$ at $\simeq$2.5$\sigma$, whereas PICO+Euclid would be able to do so at $\simeq$3.2$\sigma$.

\begin{figure*}
\centering
\includegraphics[width=0.8\linewidth]{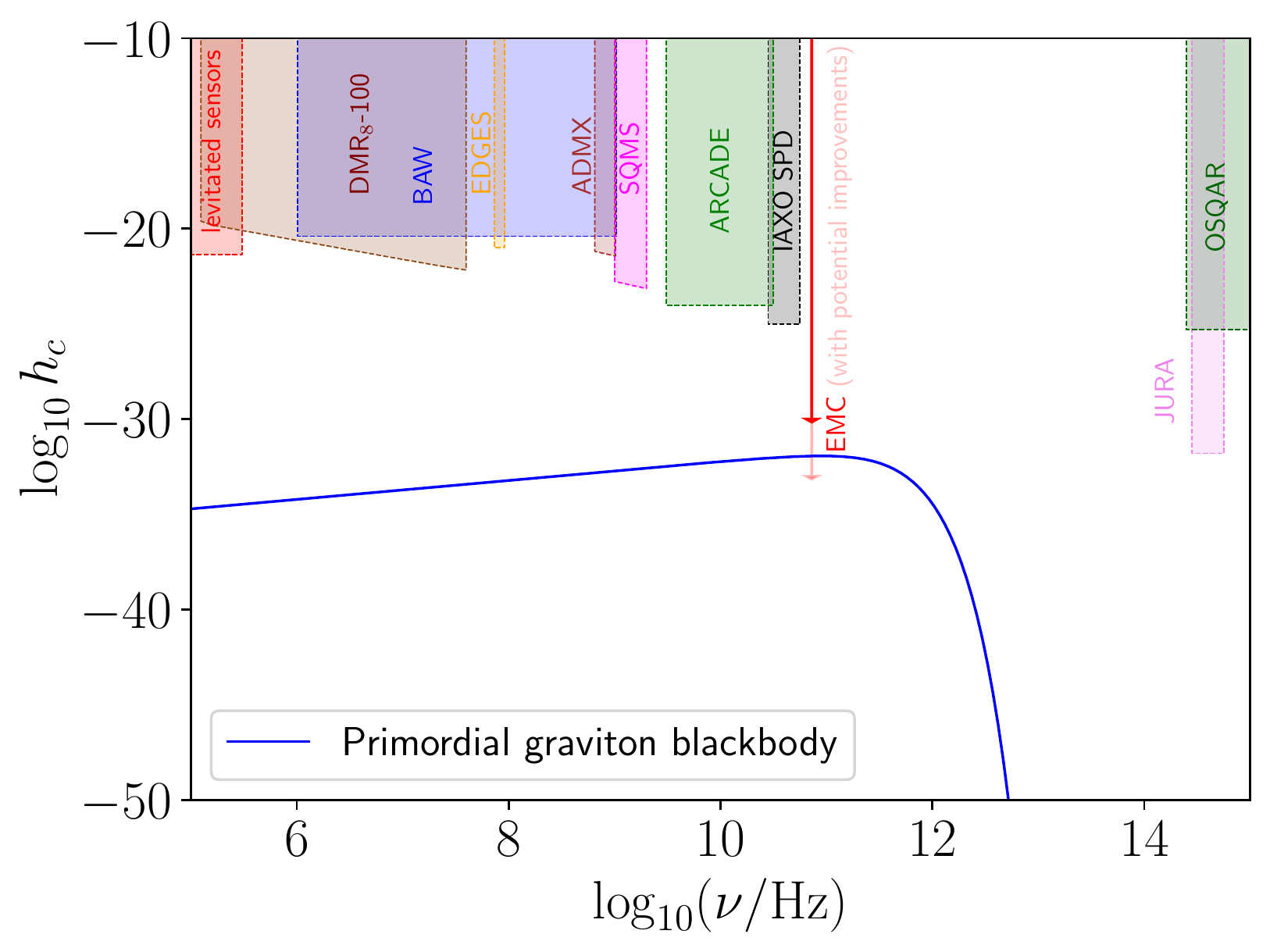}
\caption{Strain of the CGB stochastic background of high-frequency GWs, alongside the sensitivities of various detector concepts discussed in the main text. The red line (``EMC'') refers to enhanced magnetic conversion, with the more transparent extension referring to potential future technological improvements discussed in the main text.}
\label{fig:primordial_graviton_blackbody}
\end{figure*}

Should the CGB contribution to $N_{\rm eff}$ be detected, one may wonder how we know that the excess radiation density is associated to the CGB, rather than another dark radiation component. To remove this ambiguity, we consider the stochastic background of (high-frequency) gravitational waves (GWs) associated to the CGB. It is useful to think in terms of characteristic strain $h_c$, i.e.\ the dimensionless strain which would be produced due to the passing stochastic GW background (SGWB) in the arms of an interferometer with arms of equal length $L$ in the $x$ and $y$ directions, $h_c(\nu) \simeq \Delta L/L$. The characteristic CGB strain $h_g(\nu)$ is given by:
\begin{align}
h_g(\nu) = \frac{1}{\nu}\sqrt{\frac{3H_0^2}{2\pi^2}\Omega_g(\nu)} \approx 1.26 \times 10^{-27} \left ( \frac{\nu}{{\rm GHz}} \right )^{-1}\sqrt{h^2\Omega_g(\nu)}\,.
\label{eq:cgbstrain}
\end{align}
where $h^2\Omega_g(\nu)$ is the CGB spectral energy density in units of the present-day critical density:
\begin{eqnarray}
h^2\Omega_g(\nu) = \frac{15}{\pi^4}h^2\Omega_{\gamma,0} \left ( \frac{T_{g,0}}{T_{\gamma,0}} \right )^4F(x_g)\,,
\label{eq:h2omegagnu}
\end{eqnarray}
with $h$ the reduced Hubble parameter, $h^2\Omega_{\gamma,0}$ the photon density parameter today, $x_g \equiv h\nu/(k_BT_{g,0})$, and $F(x_g) \equiv x_g^4/(e^{x_g}-1)$. The CGB spectrum peaks at frequencies $\nu \approx 75\,{\rm GHz}$, making it a source of a high-frequency GWs: Fig.~\ref{fig:primordial_graviton_blackbody} shows the characteristic CGB strain alongside demonstrated or forecasted sensitivities of various detector concepts~\citep[see][]{Aggarwal:2020olq}.

Aside from optically levitated sensors~\citep{Arvanitaki:2012cn} and bulk acoustic wave (BAW) devices~\citep{Goryachev:2014yra}, all probes in Fig.~\ref{fig:primordial_graviton_blackbody} exploit the \textit{inverse Gertsenshtein effect} (IGE), whereby GWs convert to photons within a strong magnetic field~\citep{Gertsenshtein:1962ghw}. While apart from small prototypes dedicated instruments exploiting the IGE do not exist, \cite{Ito:2019wcb} and~\cite{Ejlli:2019bqj} showed how constraints on high-frequency GWs can be obtained re-interpreting data from ongoing or planned axion experiments: in Fig.~\ref{fig:primordial_graviton_blackbody} this includes ADMX, SQMS, IAXO SPD, JURA, OSQAR, and DMRadio$_8$-100~\citep{Domcke:2022rgu}. The IGE can also be exploited in strongly magnetized astrophysical environments~\citep{Chen:1994ch,Domcke:2020yzq}, recasting observations from radio telescopes such as EDGES and ARCADE. For more details on these detector concepts, see~\cite{Aggarwal:2020olq,Berlin:2021txa,Domcke:2022rgu}.

Unfortunately, as is clear from Fig.~\ref{fig:primordial_graviton_blackbody}, all these detector concepts fall short of the CGB signal by several orders of magnitude. The only promising probe is enhanced magnetic conversion (EMC), a proposal to enhance the efficiency of IGE-based magnetic conversion detectors by seeding the conversion volume with locally generated auxiliary EM fields, e.g.\ EM Gaussian beams (GBs) oscillating at the frequency of the GW signal searched for~\citep{Li:2004df,Baker:2008zzb}. Until recently, EMC appeared to be well beyond technological reach, particularly due to the requirement of a GB geometric purity at the $10^{-21}$ level to reach strain levels of $h_c \sim 10^{-30}$ at $\nu \sim {\cal O}(100)\,{\rm GHz}$.

However, \cite{Ringwald:2020ist} argued that reaching the above benchmark limit is feasible, exploiting state-of-the-art superconducting magnets utilized in near-future axion experiments to generate the required EM signal, then enhanced by a GB produced by a MW-scale $40\,{\rm GHz}$ gyrotron. While this still leaves us 2 orders of magnitude short of the CGB peak strain, realistic improvements in the development of gyrotrons, single-photon detectors (SPDs), and superconducting magnets, can bring the projected sensitivity down to $h_c \sim 10^{-32}$, sufficient to detect the CGB in our benchmark scenario. We estimate that an increase in the gyrotron available power to $\sim 100\,{\rm MW}$ (which is realistically achievable) over a stable running time of $\sim 1\,$month (which is much more challenging), alongside improvements in SPDs dark count rates to $\sim 10^{-5}\,{\rm s}^{-1}$, would result in a sensitivity to strains of order $h_c \sim 10^{-33}$, sufficient to detect our benchmark CGB. All quoted sensitivities can be further improved by increasing the reflector size, and the intensity and length of the magnets. Therefore, measuring strains as small as $h_c \sim 10^{-33}$ at $\nu \sim {\cal O}(100)\,{\rm GHz}$, and detecting the benchmark CGB, might be feasible in the not-too-far-off future.~\footnote{We recall that the quoted CGB strength assumes that the SM holds up to the Planck scale, and that the appearance of additional DoF would lower the CGB temperature and associated SGWB strength. However, even in the extremely unrealistic scenario where the number of DoF increases by an order of magnitude, the temperature of the CGB would only decrease by a factor of $\gtrsim 2$, making the SGWB signal only a factor of $\approx 5$ weaker [see the $T_{g,0}$ dependence in Eqs.~(\ref{eq:cgbstrain},\ref{eq:h2omegagnu})].}

Another interesting potential detection channel proposed very recently by~\cite{Brandenberger:2022xbu} proceeds through a parametric instability of the EM field in the presence of GWs. This would allow for conversion of high-frequency GWs to photons without the need for a strong background magnetic field. Sensitivity reach estimates for this probe, while not yet available, are worth further investigation in this context.

An important issue concerns how to distinguish the CGB from competing SGWB sources. Possible examples could be the SGWB produced during preheating~\citep{Easther:2006gt} or during oscillon formation~\citep{Zhou:2013tsa}: however, both these sources are important at lower frequencies, ${\cal O}(10^6-10^9)\,{\rm Hz}$~\citep[see][]{Aggarwal:2020olq}, and hence should not confuse the CGB detection. The CGB SGWB can also be distinguished from the SGWB produced by out-of-equilibrium gravitational excitations at reheating~\citep{Ringwald:2020ist}: the latter would not be of the blackbody form, and its strength would be orders of magnitude below the CGB as long as the reheating temperature is $T_{\rm rh} \ll M_{\rm Pl}$, which as argued earlier can be safely assumed. This highlights the importance of detecting the CGB over a range of frequencies, given the clear prediction for its frequency dependence. Within the EMC experimental setup, this can be achieved by tuning the gyrotron frequency: the output frequencies available for typical gyrotrons fall within the $\sim 20-500\,{\rm GHz}$ range, perfectly suited to probe the CGB spectrum around its peak frequency. A similar tuning procedure should also be possible for the GW-photon parametric instability probe.

A caveat to our findings is the assumption of a pure blackbody spectrum for primordial gravitons. This is likely to be an approximation at best, particularly at low frequencies, whose modes would have been super-horizon at the Planck time. However, in the absence of detailed knowledge regarding the underlying theory of quantum gravity, this is among the most conservative assumptions we can make~\citep[note that the same approximation has been made in several earlier works discussing primordial gravitons, e.g.][]{Zhao:2009pt,Giovannini:2019oii}. Moreover, what is important for our results is the high-frequency tail of the CGB spectrum, where our assumption is likely to be far more realistic. Overall, it remains true that finding any trace of a GW background of the estimated amplitude at the estimated frequency will rule out the standard inflationary scenario.

\section{Alternatives to inflation}
\label{sec:alternatives}

Our previous discussion raises the question of whether an unambiguous CGB detection would also spell trouble for alternative paradigms, where density perturbations are produced during a non-inflationary phase. While the answer to this question is highly model-dependent, we wish to provide a brief qualitative assessment limited to two well-motivated paradigms: bouncing cosmologies and emergent scenarios.

Within bouncing cosmologies, the challenge is to produce a thermal CGB in first place. This is hard to achieve during the contracting phase, when the characteristic energy scale is typically $\Lambda_c \ll M_{\rm Pl}$~\citep[e.g.][]{Brandenberger:2016vhg}. Another possibility is one where a relatively long bouncing phase with energy density around the Planck scale occurs between the initial contracting phase and the later hBB expansion~\citep[e.g.][]{Cai:2014bea}, in which case a thermal CGB would be generated and would survive the phase transition between the bouncing and expanding phases.

In emergent scenarios, the Universe emerges from an initial high density state with matter in global thermal equilibrium, and producing the CGB is far less unlikely. A particularly well-studied emergent scenario is the string gas proposal of~\cite{Brandenberger:1988aj}, where the Universe originates from a quasi-static Hagedorn phase of a string gas at temperature close to the Hagedorn temperature, before a T-dual symmetry breaking-driven phase transition connects to the hBB expansion. On general grounds, the energy density in the emergent phase is close to the Planck density, making it likely for gravitons to be in thermal equilibrium and therefore for a CGB to be generated. 

However, the initial state in string gas cosmology is not a thermal state of particles but of strings, giving a different scaling of thermodynamical quantities. It is therefore unlikely that the string gas CGB takes the blackbody form, although it is in principle possible that its spectral energy density may be higher than our benchmark CGB, enhancing detection prospects. Fully exploring these points requires a dedicated study, going beyond the scope of our work.

\section{Conclusions}
\label{sec:conclusions}

Despite its enormous success, recent debates around the inflationary paradigm raise the question of how to \textit{model-independently} discriminate it from competing scenarios for the production of primordial density perturbations. In this \textit{Letter}, we have argued that a detection of the Cosmic Graviton Background (CGB), the left-over graviton radiation from the Planck era, would rule out the inflationary paradigm, as realistic inflationary models dilute the CGB to an unobservable level. Assuming the validity of the SM up to the Planck scale, the CGB contribution to the effective number of relativistic species $\Delta N_{{\rm eff},g} \approx 0.054$ is well within the reach of next-generation cosmological probes, whereas detecting the associated stochastic background of high-frequency GWs in the $\nu \sim {\cal O}(100)\,{\rm GHz}$ range is challenging but potentially feasible. We also argued that the CGB may be detectable within well-motivated alternatives to inflation such as bouncing and emergent scenarios. We hope that this work will spur further investigation into the possibility of model-independently confirming or ruling out the inflationary paradigm with upcoming observations~\citep[for similar endeavors see e.g.][]{Chen:2018cgg}.

\section*{Acknowledgements}

We are grateful to Robert Brandenberger, Massimo Giovannini, Will Kinney, Nick Rodd, and Luca Visinelli for useful discussions and suggestions. S.V. is partially supported by the Isaac Newton Trust and the Kavli Foundation through a Newton-Kavli Fellowship, by a grant from the Foundation Blanceflor Boncompagni Ludovisi, n\'{e}e Bildt, and by a College Research Associateship at Homerton College, University of Cambridge. A.L. is partially supported by the Black Hole Initiative at Harvard University, which is funded by grants from the John Templeton Foundation and the Gordon and Betty Moore Foundation.

\bibliography{Primordial_graviton_background}{}
\bibliographystyle{aasjournal}

\label{lastpage}
\end{document}